\documentclass[aps,prl,showpacs,twocolumn]{revtex4}

\usepackage{graphicx}
\usepackage{amsmath}
\usepackage{amsfonts}
\usepackage{color}
\usepackage{epsfig}

\usepackage{appendix}

\begin{document}

\title{Minimal Entangled States and Modular Matrix for Fractional Quantum Hall Effect in Topological Flat Bands}

\author{W. Zhu$^1$, D. N. Sheng$^1$, F. D. M. Haldane$^2$}
\affiliation{$^1$Department of Physics and Astronomy, California State University, Northridge, California 91330, USA}
\affiliation{$^2$Department of Physics, Princeton University, Princeton, NJ 08544, USA}

\begin{abstract}
We perform an exact diagonalization study of the topological order in
topological flat band  models through calculating entanglement entropy and spectra of low energy states.
We identify multiple independent minimal entangled states, which form a set of orthogonal basis states
for the groundstate manifold.
We extract the modular transformation matrices $\mathcal{S}$ ($\mathcal{U}$) which
contains the information of  mutual (self) statistics,  quantum dimensions and fusion rule of quasiparticles.
Moreover, we demonstrate that these matrices are robust and universal in the whole topological phase
against different perturbations until the quantum phase transition takes place.
\end{abstract}


\maketitle

\textit{Introduction.---}
The fractional quantum Hall (FQH) state is the best-known many-body state
with topological order discovered in 2D electron systems
under strong magnetic field.
The most striking features of FQH state are the topological ground state
degeneracy on torus and the emerging quasiparticles obeying fractional statistics
\cite{Laughlin1983,Wen1989}.
Recently, it has been demonstrated that FQH states can also be realized in
various topological flat-band (TFB) models without Landau levels
\cite{Haldane1988,Tang,Neupert,Sun,
DNSheng2011,YFWang2011,Bernevig2011,Bernevig2012,YFWang2012,Bernevig2013,
XLQi2012,Moller,DXiao,Shankar,Venderbos,ZLiu,Kapit,Shuo,Sondhi}.
In such an interacting system, an explicit demonstration of topological order and
quasiparticle statistics is still highly desired, which has
attracted lots of recent interests
\cite{Wen1990,Kitaev,Levin,Haldane2008,Lauchli2010a,HCJiang,YZhang2012,Vidal,Pollmann,Tu}.

Entanglement measurements such as
topological entanglement entropy (TEE) \cite{Kitaev, Levin}
and entanglement spectrum \cite{Haldane2008}
have been identified as powerful tools for detecting topological properties
of many-body quantum states. Insightfully,
Zhang \textit{et al.} proposed to extract modular matrix
through the entanglement measurement \cite{YZhang2012},
which encodes the complete information
of the topological order including quasiparticles quantum dimension and
statistics as first described by Wen \cite{Wen1990}.
Based on the model  wavefunctions for toric code and chiral spin liquid states,
they demonstrated that the transformation between
the minimal entangled states (MESs)
along two interwinding partition directions
gives rise to modular matrices.
The new route to extract modular matrix through MESs improves the practical
implementation  for strongly interacting systems
as such information is accessible through larger system density matrix renormalization group (DMRG) calculations demonstrated for  bosonic FQH state $\nu=1/2$ in TFB \cite{Vidal} and fermionic FQH states with magnetic field\cite{Pollmann}.
However, it remains difficult to access multiple low energy states in DMRG in a controlled way,
when there are coupling between different topological sectors induced
by interaction or when the groundstates have higher degeneracy,
which will be the focuses of our exact diagonalization (ED) study.

In this letter, we present an ED calculation
for the TFB model and map out the entanglement entropy profile for
superposition states of the near degenerating
groundstates. We demonstrate that there are
the same number of the  MESs as the ground state degeneracy for FQH phase
on a torus,
which form the orthogonal and complete basis states
for modular transformation. Through locating the MESs along two interwinding
partition directions, we extract  the modular matrices $\mathcal{S}$ and $\mathcal{U}$ containing
generalized statistics of quasiparticles,
which unambiguously demonstrate the fractional
quasiparticle statistics in such systems
for $\nu=1/2, \nu=1/4$ (bosons),
and $\nu=1/3$ (fermions) FQH states, respectively.
We also analyze the entanglement spectra and obtain  TEE  from
the difference of the maximum and minimum of entanglement entropies
of these superposition states.
Furthermore, we study the quantum phase transition from FQH phase to
the topological trivial phase driven by the disorder scattering or attractive
anisotropic interaction. Significantly,
the extracted modular matrices
 remain to be universal containing the same  quasi-particle fractional statistics information as 
theoretical ones for the model FQH states in the whole topological phase until
the quantum phase transition takes place.
This is distinctly different from following the Berry phase of
the ground states, where only the sum  of the total
Chern number remains invariant\cite{DNSheng2003} due to the lifting of the degeneracy by perturbations for any finite size systems.

We study the Haldane model~\cite{Haldane1988} on the honeycomb (HC) lattice:
\begin{eqnarray}\label{hamilton}
H_{\rm HC}&=& -t^{\prime}\sum_{\langle\langle\mathbf{r}\mathbf{r}^{
\prime}\rangle\rangle}
\left[c^{\dagger}_{\mathbf{r}^{ \prime}}c_{\mathbf{r}}\exp\left(i\phi_{\mathbf{r}^{ \prime}\mathbf{r}}\right)+\textrm{H.c.}\right]\nonumber\\
&-&t\sum_{\langle\mathbf{r}\mathbf{r}^{ \prime}\rangle}
\left[c^{\dagger}_{\mathbf{r}^{\prime}}c_{\mathbf{r}}+\textrm{H.c.}\right]
-t^{\prime\prime}\sum_{\langle\langle\langle\mathbf{r}\mathbf{r}^{
\prime}\rangle\rangle\rangle}
\left[c^{\dagger}_{\mathbf{r}^{\prime}}c_{\mathbf{r}}+\textrm{H.c.}\right]\nonumber \\
&+&V_1\sum_{\langle\mathbf{r}\mathbf{r}^{\prime}\rangle}n_{\mathbf{r}}n_{\mathbf{r}^{\prime}}
+\sum_{\mathbf{r}} \epsilon_{\mathbf{r}} c^{\dagger}_{\mathbf{r}} c_{\mathbf{r}}
\label{e.1}
\end{eqnarray}
where $c^{\dagger}_{\mathbf{r}}$ 
creates a hard-core boson (or fermion) at site
$\mathbf{r}$,  $n_{\mathbf{r}}=c^{\dagger}_{\mathbf{r}}c_{\mathbf{r}}$
is the boson (or fermion) number operator.
$\langle\dots\rangle$, $\langle\langle\dots\rangle\rangle$ and
$\langle\langle\langle\dots\rangle\rangle\rangle$ denote the nearest-neighbor (NN), the
next-nearest-neighbor (NNN) and the next-next-nearest-neighbor (NNNN) pairs of sites,
and $V_1$ is the NN interaction. The last term models the Anderson on-site disorder $\epsilon_{\mathbf{r}}$ randomly distributed in $[-W,W]$.
On the HC lattice, we select the parameters $t=1$, $t^{\prime}=0.60$,
$t^{\prime\prime}=-0.58$ and the magnitude of the
hopping phase $\phi=0.4\pi$, which lead to a topological
flat-band with flatness ratio about $50$ \cite{YFWang2011}.
We consider
a finite system of $N_1\times N_2$ unit cells (total number of sites
$N_s=2\times N_1\times N_2$) with periodic boundary conditions.
The filling factor is $\nu=N_p/(N_1N_2)$, where $N_p$ is the number of particles.
We denote the momentum vector $(2\pi k_1/N_1,2\pi k_2/N_2)$ with
$(k_1,k_2)$ as integer quantum numbers.

\begin{figure}[t]
 \begin{minipage}{0.55\linewidth}
  \includegraphics[width=2.5in]{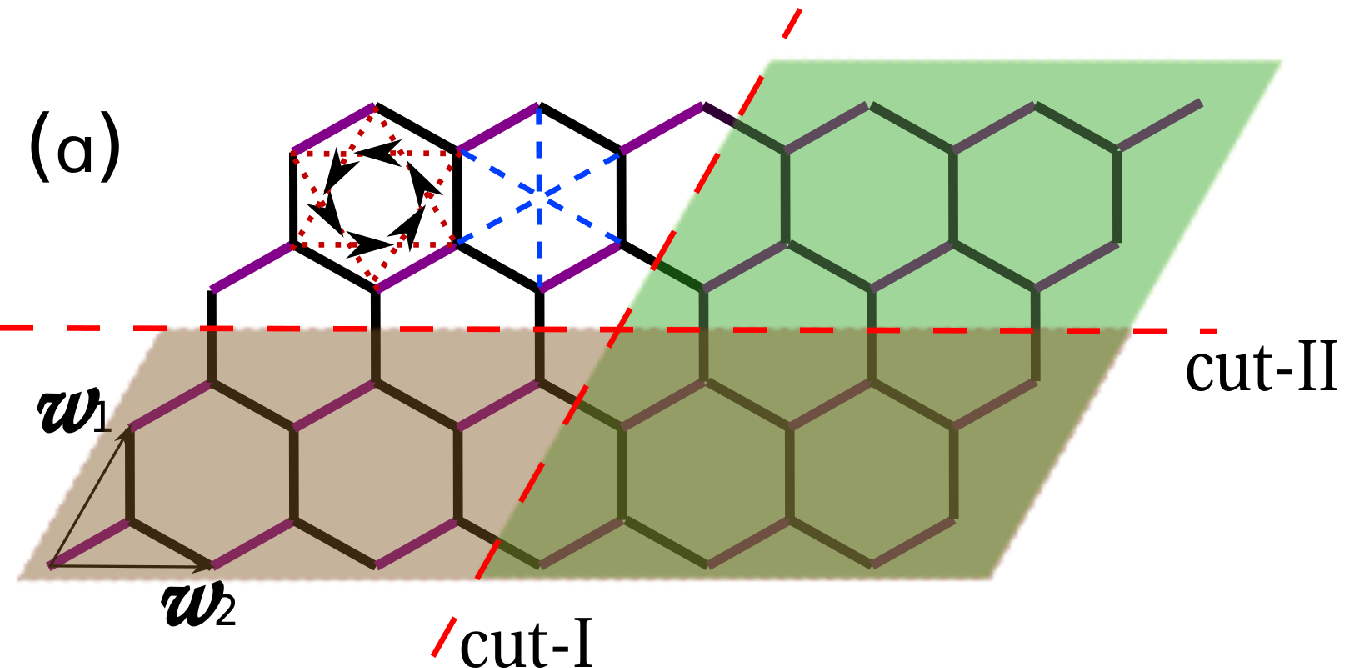}
 \end{minipage}
 \begin{minipage}{0.49\linewidth}
  \includegraphics[width=1.6in]{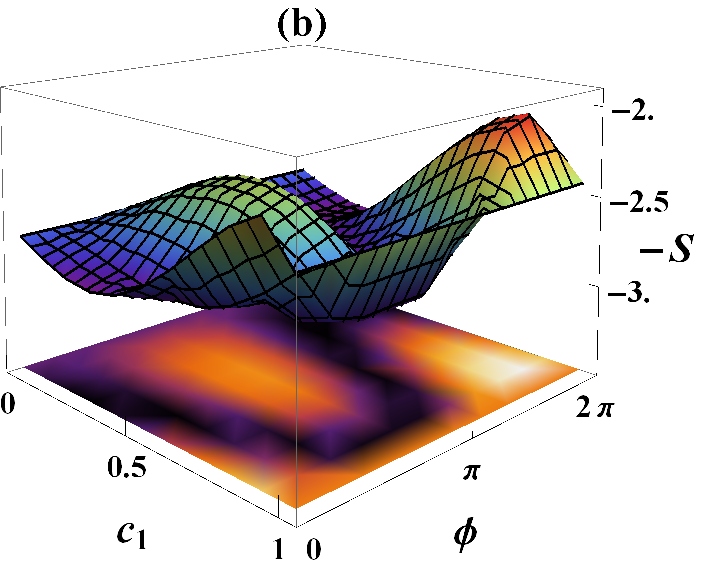}
 \end{minipage}
 \begin{minipage}{0.49\linewidth}
  \includegraphics[width=1.6in]{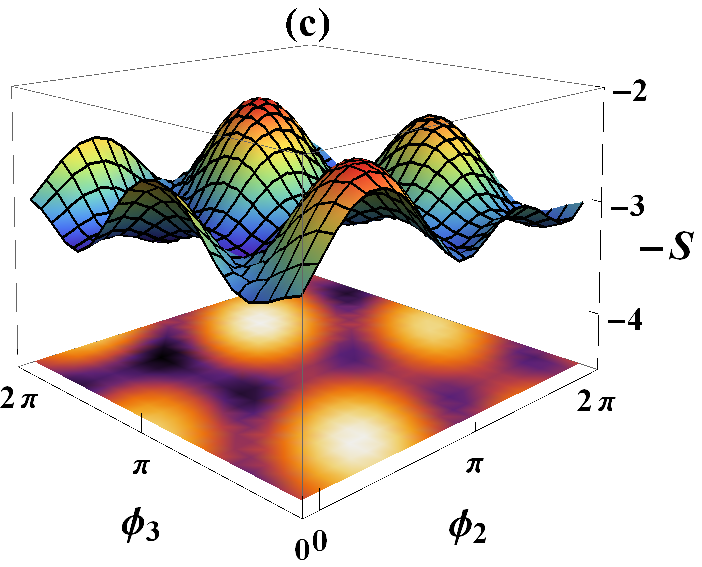}
 \end{minipage}
 \caption{(Color online) (a) Haldane model on $2 \times 4\times 6$ HC lattice with lattice vectors $\vec{w}_1,\vec{w}_2$. The arrow directions on red dotted lines present the signs of the phases $\pm\phi$ in the NNN hopping terms. The NNNN hoppings are represented by the blue dashed lines. The two ways to bipartition the system along dashed lines are labeled as cut-I and cut-II. (b) Surface and contour plots of Renyi $n=2$ entanglement entropy ($-S$) of wavefunction $|\Phi(c_1, \phi)>$ on $2\times 4\times 4$ HC lattice filled with $8$ hard-core bosons.
 (c) The entropy ($-S$) of wavefunction $|\Psi_{(c_1,c_2,\phi_2,\phi_3)}>$ on $2\times 4\times 6$ HC lattice  with $8$ interacting fermions.
}
\label{fig:hc:3D}
\end{figure}

The entanglement entropy is defined by partitioning the full system
into two subsystems A and B. Tracing out the subsystem $B$, one can obtain
the reduced density matrix of subsystem $A$:
$\rho_A=tr_B|\Phi><\Phi|$, where $|\Phi>$ is the many-body state of the full system.
The Renyi $n=2$ entanglement entropy is defined as:
$S=-\log tr\rho^2_A$. Here we focus on
two noncontractible bipartitions on a torus,
as shown in Fig.~\ref{fig:hc:3D}(a)
as cut-I and cut-II, respectively.

\textit{Multiple MESs as superpositions of near degenerating groundstates.---}
In TFB lattice model, it has been identified that there
are $m$ near degenerating  groundstates at filling factor $\nu=1/m$\cite{YFWang2011,DNSheng2011}
when the interacting system realizes  a FQH phase.
Let us first consider a $2\times 4\times 4$ HC lattice
filled with hard-core bosons at half-filling \cite{YFWang2011}.
We set the NN interaction to be zero since hard-core bosons are
intrinsically interacting.
From ED calculation, we
find the two groundstates
 $|\xi_1>$ and $|\xi_2>$ both
 in the same momentum sector $(k_1,k_2)=(0,0)$.
This is the general case as long as the two system lengths $N_1$ and $N_2$
are factors of the particle number $N_p$.
Now we form the general superposition
state as,
\begin{equation*} 
|\Phi_{c_1,\phi}>=c_1|\xi_1>+c_2e^{i\phi}|\xi_2>
\end{equation*}
where $c_1$ and $\phi$ are the real parameter and the relative phase
of the state respectively, while $c_2=\sqrt{1-c_1^2}$.
For each state $|\Phi_{c_1, \phi}>$,  we construct the reduced density
matrix and obtain the corresponding entanglement entropy.
In Fig.~\ref{fig:hc:3D}(b), we draw the  $-S$ in the surface and contour plots
so that the peaks in entropy show up clearly representing the minimums  of $S$.
We  identify two peak structures in $(c_1,\phi)$ parameter space
corresponding to two independent MESs:
\begin{eqnarray}
\label{MES:hc:boson}
  |\Xi^I_1> &=& 0.892|\xi_1>+0.451 e^{i1.74\pi}|\xi_2> \nonumber\\
  |\Xi^I_2> &=& 0.455|\xi_1>+0.890 e^{i0.74\pi}|\xi_2>
\end{eqnarray}
The  minimal entropies at the two peaks are
different with $S=2.044$ and $2.388$ respectively, indicating the
finite size effect.
However, we find the relative phase difference between the two MESs is $\phi(1)-\phi(2)=\pi$ and
consequently the two MESs are approximately orthogonal to each other: $|<\Xi_1|\Xi_2>|\approx 0.005$.
Due to the $\pi/3$ rotation symmetry in the $2\times 4\times 4$ system,
the MESs along cut-II $|\Xi^{II}_i>$ are related to
$|\Xi^{I}_i>$ as $|\Xi^{II}_i>=R_{\frac{\pi}{3}}|\Xi^{I}_i>,i=1,2$,
where $R_{\frac{\pi}{3}}$ is the $\pi/3$ rotation operator.

Now we further examine  the relation between the MESs and the
degeneracy of the ground state manifold
by studying the TFB model filled with fermions at $\nu=1/3$.
We consider a $2\times 4\times 6$ HC lattice with $8$
fermions with repulsive NN $V_1=1$
to stabilize the FQH  phase \cite{DNSheng2011}.
In the ED study, we find three
quasi-degenerating groundstates  $|\xi_j>$, (with $j=1,2,3$)
in momentum sectors $(k_1,k_2)=(0,0),(0,2),(0,4)$, respectively.
We search for the superposition
states in the space of the groundstate manifold with minimal entropy
using the following general wavefunctions:
\begin{equation*} 
|\Psi_{(c_1,c_2,\phi_2,\phi_3)}>=c_1|\xi_1>+c_2e^{i\phi_2}|\xi_2>+c_3e^{i\phi_3}|\xi_3>
\end{equation*}
where $c_1,c_2$ are real parameters
and $\phi_2$, $\phi_3$ are relative phases for the state, while
$c_3$ can be obtained using normalization condition. 
For the bipartition along cut-I, we observe two key points:
1) We can locate three global minimal entropy states in the given parameter space, which always occur when $|c_1|=|c_2|=|c_3|\approx 1/\sqrt{3}$;
2) The relative phases of two different  MESs $i$ and $j$
 satisfy:
$\phi_m(i)-\phi_m(j)\approx \pm\frac{2\pi}{3}$, for $m=2,3$.
In Fig.~\ref{fig:hc:3D}(c), we show the surface and
contour plots of the entropy of the state $|\Psi_{c_1,c_2,\phi_2,\phi_3}>$
as functions of $\phi_2$ and $\phi_3$
while other parameters are fixed  at $c_1=c_2=c_3=\frac{1}{\sqrt{3}}$
so that MESs occur with varying the relative phases.
The three MESs are determined as:
\begin{equation}
\label{MES:fermi:x}
  |\Xi^{I}_i> = (|\xi_1> +e^{i\phi_2(i)}|\xi_2> + e^{i\phi_3(i)}|\xi_3>)/\sqrt{3} \,\,\,\,\,\,
\end{equation}
with state index $i=1,2,3$. We find $(\phi_2(1),\phi_3(1))=(0.546\pi,0.286\pi)$, $(\phi_2(2),\phi_3(2))=(1.220\pi,1.620\pi)$, $(\phi_2(3),\phi_3(3))=(1.854\pi,0.930\pi)$,
corresponding to minimum entropies  $S=2.309,2.309,2.464$, respectively.
Very importantly,  the three MESs we found are nearly orthogonal to each other: $|<\Xi^{I}_1|\Xi^{I}_2>|\approx 0.007$, $|<\Xi^{I}_3|\Xi^{I}_1>|\approx 0.030$ and $|<\Xi^{I}_3|\Xi^{I}_2>|\approx 0.025$,  which is the necessary condition for these states to form the basis states for modular transformation.
The small overlap is a finite size effect as the MESs become  the true
ground states only in the thermodynamic limit.
Since there is no  rotation symmetry in $2\times 4\times 6$ lattice,
we separately locate three MESs in the parameter space
for the partition along cut-II.
Here, we find that  each  groundstate $|\xi_i>$ is indeed the MES
\begin{equation}
\label{MES:fermi:y}
  |\Xi^{II}_i> = |\xi_i> ,\,\,\, i=1,2,3. 
\end{equation}
In general, if the groundstates have different
momentum along the entanglement cut direction, these states
are eigenstates with definite number of quasiparticles,
and thus any form of mixing   will increase  the  entropy of the state.

We have also studied the TFB model on checkboard (CB) lattice \cite{Sun,DNSheng2011}
and obtained similar results. Interestingly,  we also
identified a four fold degenerating MESs at $\nu=1/4$ filling
corresponding to  a $\nu=1/4$ FQH\cite{YFWang2011,supplement}.

\begin{figure}[b]
\centering
\includegraphics[width=0.49\textwidth]{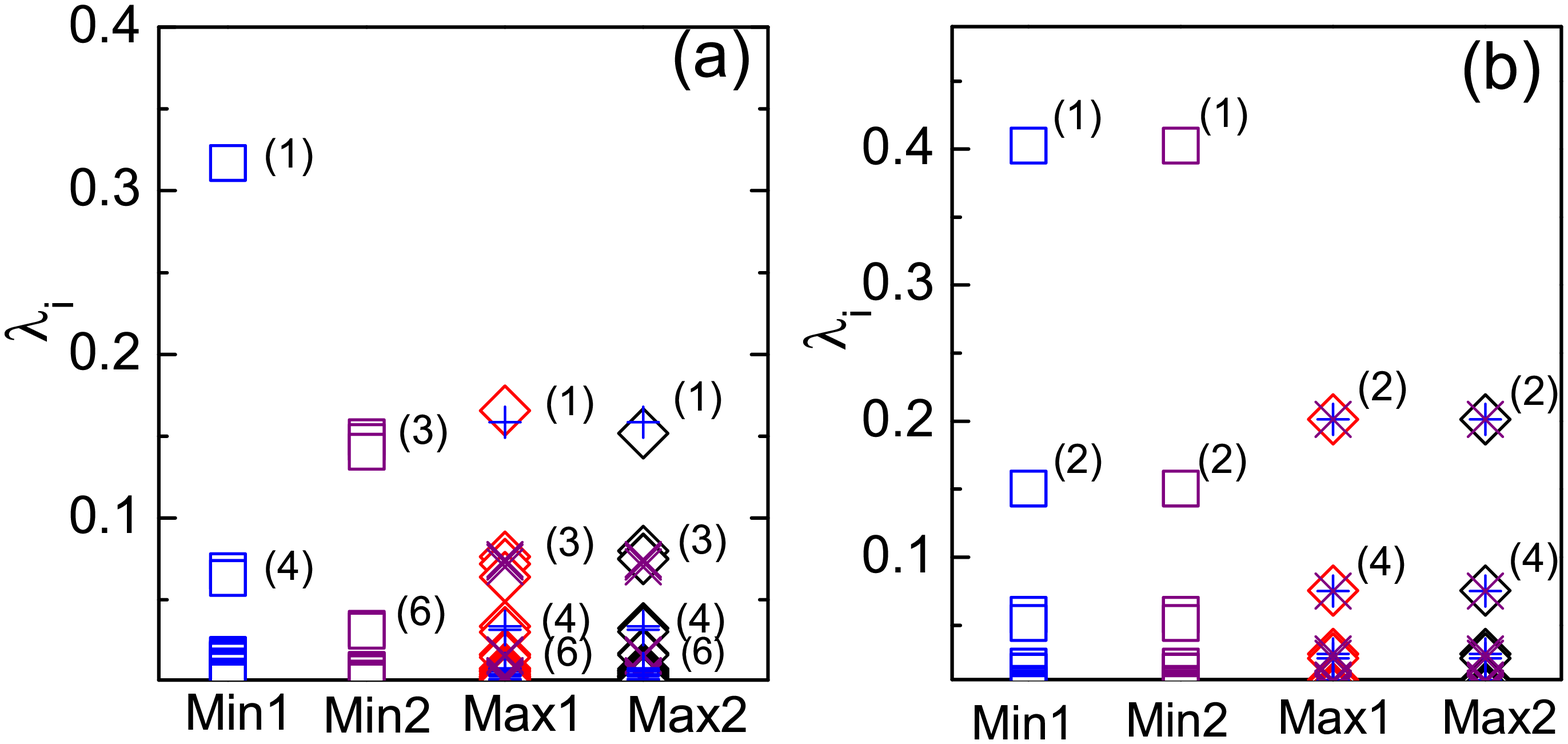}
\caption{(Color online) Eigenvalues $\lambda_i$ of the reduced density matrix
of two MESs (square) and two maximal entangled states (diamond) for
(a) $2\times 4\times 4$ and (b) $2\times 3 \times 6$ HC lattice with hard-core bosons at $\nu=1/2$.
The number near the dots shows the degeneracy.
Crossed blue (purple) dots stand for the combination of $\{\lambda^{min1(2)}_i\}$
as described in the text.}
\label{fig:es:hc}
\end{figure}

\tabcolsep=5pt
\tabcolsep=5pt
\begin{table}[t]
 \centering
 \caption{The comparison of the calculated TEE $\gamma_{cal}=S_{max}-S_{min}$
 (for  cut-I)
and the theoretical values $\gamma_{Th}=\ln m$
for $1/m$ Laughlin states. HB and FM denote hard-core boson and fermion
systems, respectively. HC and CB represent Honeycomb and checkboard lattices.
'Y(N)' means the groundstates have the same (different) momentum.}\label{table:gamma}
 \begin{tabular}{lcccc}
  \hline
  \hline
  system & lattice size & GSM & $\gamma_{cal}$ & $\gamma_{cal}/\gamma_{Th}$ \\
  \hline
  HB on HC $\nu=1/2$ & $2\times 4\times 4$ & Y & 0.849 & 1.232 \\
  HB on HC $\nu=1/2$ & $2\times 3\times 6$ & N & 0.693 & 0.999 \\
  FM on HC $\nu=1/3$ & $2\times 4\times 6$ & N & 1.125 & 1.024 \\
  HB on CB $\nu=1/2$ & $2\times 4\times 4$ & Y & 0.774 & 1.117 \\
  HB on CB $\nu=1/2$ & $2\times 3\times 6$ & N & 0.693 & 0.999 \\
  FM on CB $\nu=1/3$ & $2\times 4\times 6$ & N & 1.128 & 1.026 \\
  \hline
  \hline
 \end{tabular}
\end{table}

\textit{Modular transformation matrix based on MESs.---}
The generalized quasiparticle statistics of a topological ordered state
is captured by the modular matrix $\mathcal{S}$ and $\mathcal{U}$  as first proposed by Wen \cite{Wen1990}.
$\mathcal{S}$ ($\mathcal{U}$) determines the mutual (self) statistics of the different quasiparticles
as well as the quantum dimension and fusion rules of quasiparticles \cite{SDong,ZHWang,Fendley}.
In general, the relationship between the modular matrices and MESs
is $<\Xi^{II}|\Xi^{I}>=\mathcal{U}^{n}\mathcal{S}^{l}\mathcal{U}^{m}$,
where $n,m,l$ are integers determined by specific modular transformation on a lattice \cite{YZhang2012}.

Specially, for a $2\times 4\times 4$ HC lattice filled with hard-core bosons, the
$\pi/3$ rotation symmetry leads to the overlap $<\Xi^{II}|\Xi^{I}>=\mathcal{U}\mathcal{S}^{-1}$ \cite{YZhang2012,Vidal}.
Thus by computing the overlap using states from Eq.~\ref{MES:hc:boson},  we obtain,
\begin{eqnarray*}
\label{result:hc:boson}
\mathcal{S} \approx 0.722
\left(\begin{array}{cc}
1 & 0.957 \\
0.957 & -1
\end{array}\right)  ,\,\,\,\,
\mathcal{U} \approx e^{-i\frac{2\pi}{24}0.921}
\left(\begin{array}{cc}
1 & 0 \\
0 & 0.999i
\end{array}\right)
\end{eqnarray*}
which are nearly identical to the theoretical ones\cite{SDong,ZHWang,Fendley}
for the model FQH state:
$\mathcal{S} = \frac{1}{\sqrt{2}}\left(\begin{array}{cc}
1 & 1 \\
1 & -1 \end{array}\right)$,
$\mathcal{U} = e^{-i\frac{2\pi}{24}1}
\left(\begin{array}{cc}
1 & 0 \\
0 & i \end{array}\right)$.
From $\mathcal{S}_{i1}=d_i/D$, we determine
the quantum dimension for two type of quasiparticles as $d_1=1$,$d_2\approx 0.957$
and total quantum dimension $D\approx 1.385$ (close to $\sqrt{2}$).
$\mathcal{S}_{1i}>0$ show that one quasiparticle as a boson
while $\mathcal{S}_{22}<0$ indicates another quasiparticle
acquires a $\pi$ phase encircling themselves.
Combined with the topological spin $\theta_2\approx i$ from $\mathcal{U}_{22}$, we identify
that these quasiparticles are semions \cite{ZHWang}.
For $2\times 4\times 6$ HC lattice filled with interacting fermions,
the overlap between Eq.~\ref{MES:fermi:x} and Eq.~\ref{MES:fermi:y}
gives $<\Xi^{II}|\Xi^{I}>=\mathcal{S}$ \cite{YZhang2012}:
\begin{equation}
\label{result:hc:fermi}
\mathcal{S}\approx
\frac{1}{\sqrt{3}}\left(
       \begin{array}{ccc}
        1 & 1 & 1 \\
        1 & e^{i2\pi\times 0.337} & e^{i2\pi\times 0.683} \\
        1 & e^{i2\pi\times 0.667} & e^{i2\pi\times 0.344}
       \end{array}
     \right)
\end{equation}
The obtained result is close to the analytic prediction \cite{Wen1990,ZHWang}:
$\mathcal{S}=
\frac{1}{\sqrt{3}}\left(
       \begin{array}{ccc}
        1 & 1 & 1 \\
        1 & \omega & \omega^{2} \\
        1 & \omega^{2} & \omega
       \end{array}
     \right)$, where $\omega=e^{i\frac{2\pi}{3}}$.
The extracted mutual statistics between quasiparticles reflects the $Z_3$ statistics.
Within the same route, we also obtain modular matrix for $\nu=1/4$ FQH states
on CB\cite{supplement}:
\begin{equation*}
\mathcal{S}\approx
\frac{1}{2}\left(
       \begin{array}{cccc}
        1 & 1 & 1 & 1\\
        1 & -i & -1 & i\\
        1 & -1 & 1 & -1\\
        1 & i & -1 & -i
       \end{array}
     \right)+10^{-2}e^{i0.49\pi}\left(
            \begin{array}{cccc}
             0 & 0 & 0 & 0\\
             0 & 1 & 1 & 0\\
             0 & 0 & 0 & 0\\
             0 & 1 & 1 & 0
            \end{array}
          \right)
\end{equation*}
, which is nearly the same as  the one representing  $Z_4$ statistics: $\mathcal{S}_{nn'}=\frac{1}{2}e^{-i\frac{2\pi nn'}{4}}$.

\begin{figure*}[t]
\begin{center}
\begin{minipage}{0.16\textwidth}
\includegraphics[width=1.3in]{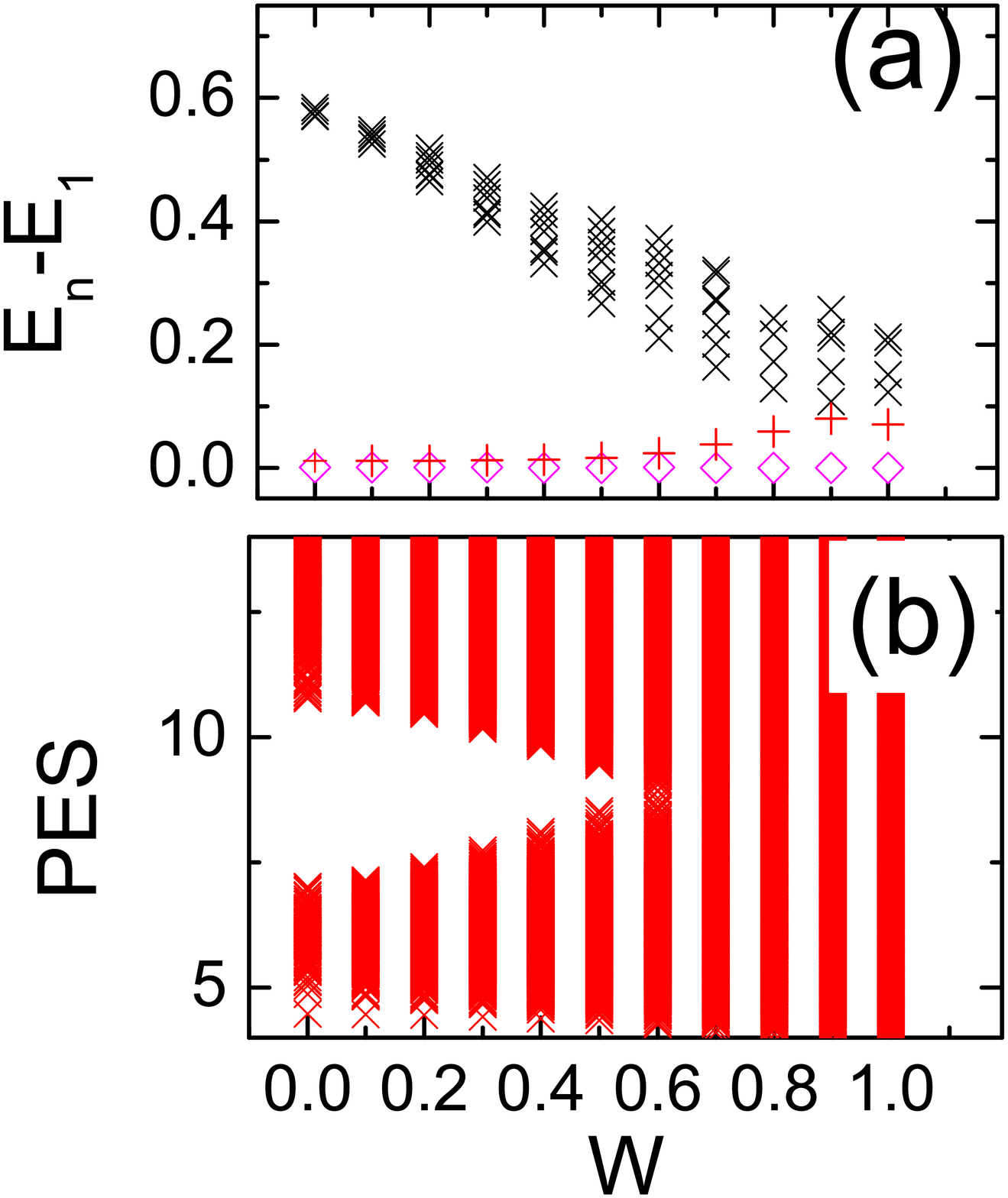}
\end{minipage}
\begin{minipage}{0.208\textwidth}
\includegraphics[width=1.5in]{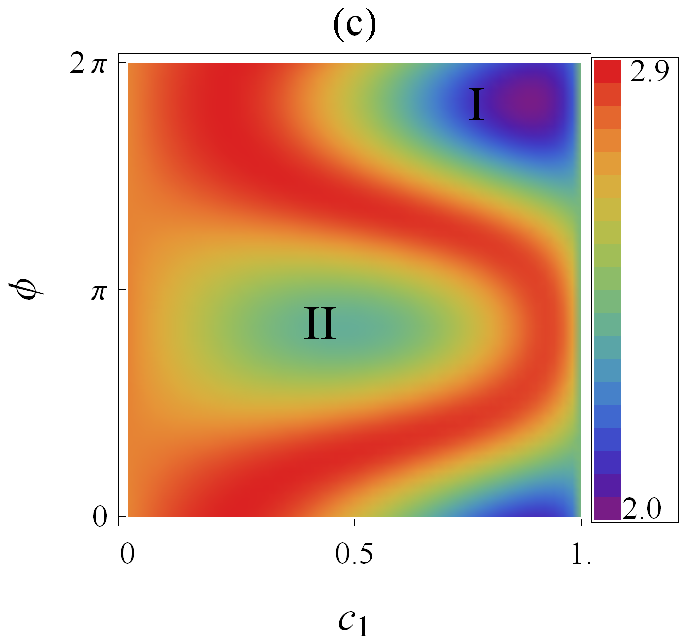}
\end{minipage}%
\begin{minipage}{0.205\textwidth}
\includegraphics[width=1.5in]{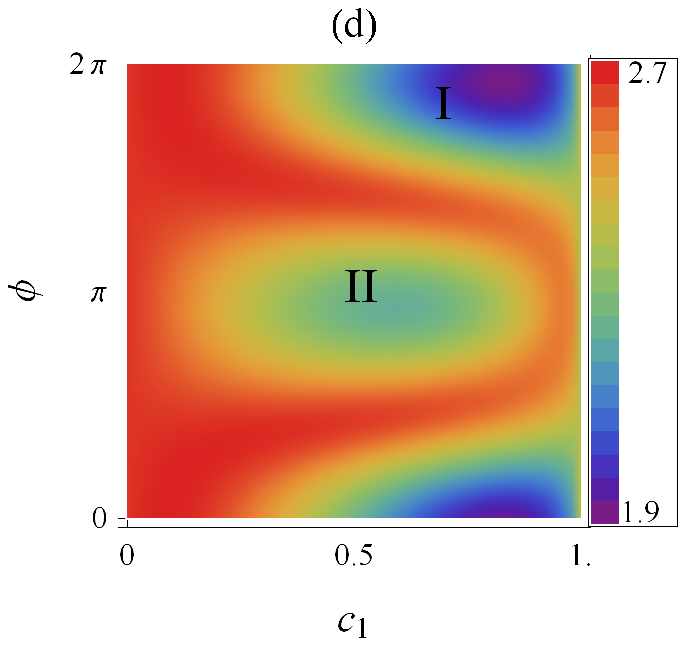}
\end{minipage}
\begin{minipage}{0.205\textwidth}
\includegraphics[width=1.5in]{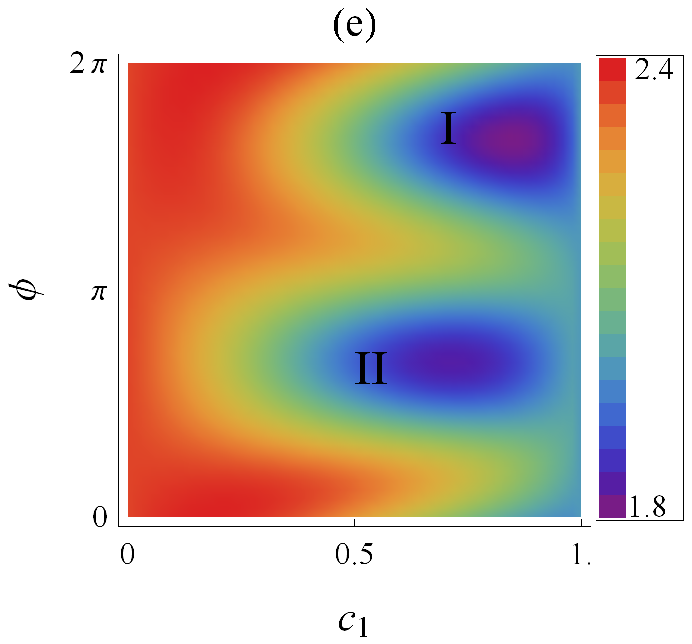}
\end{minipage}
\begin{minipage}{0.205\textwidth}
\includegraphics[width=1.5in]{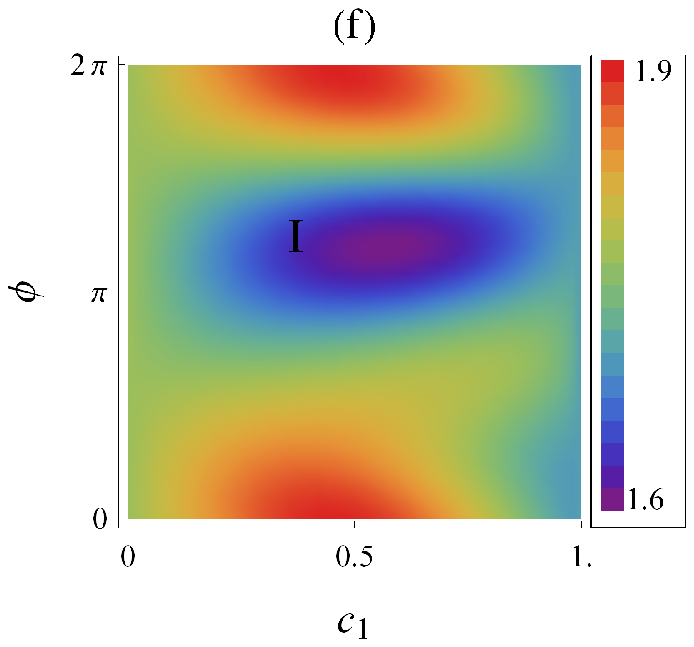}
\end{minipage}
\end{center}
\begin{center}
\begin{minipage}{0.16\textwidth}
\includegraphics[width=1.3in]{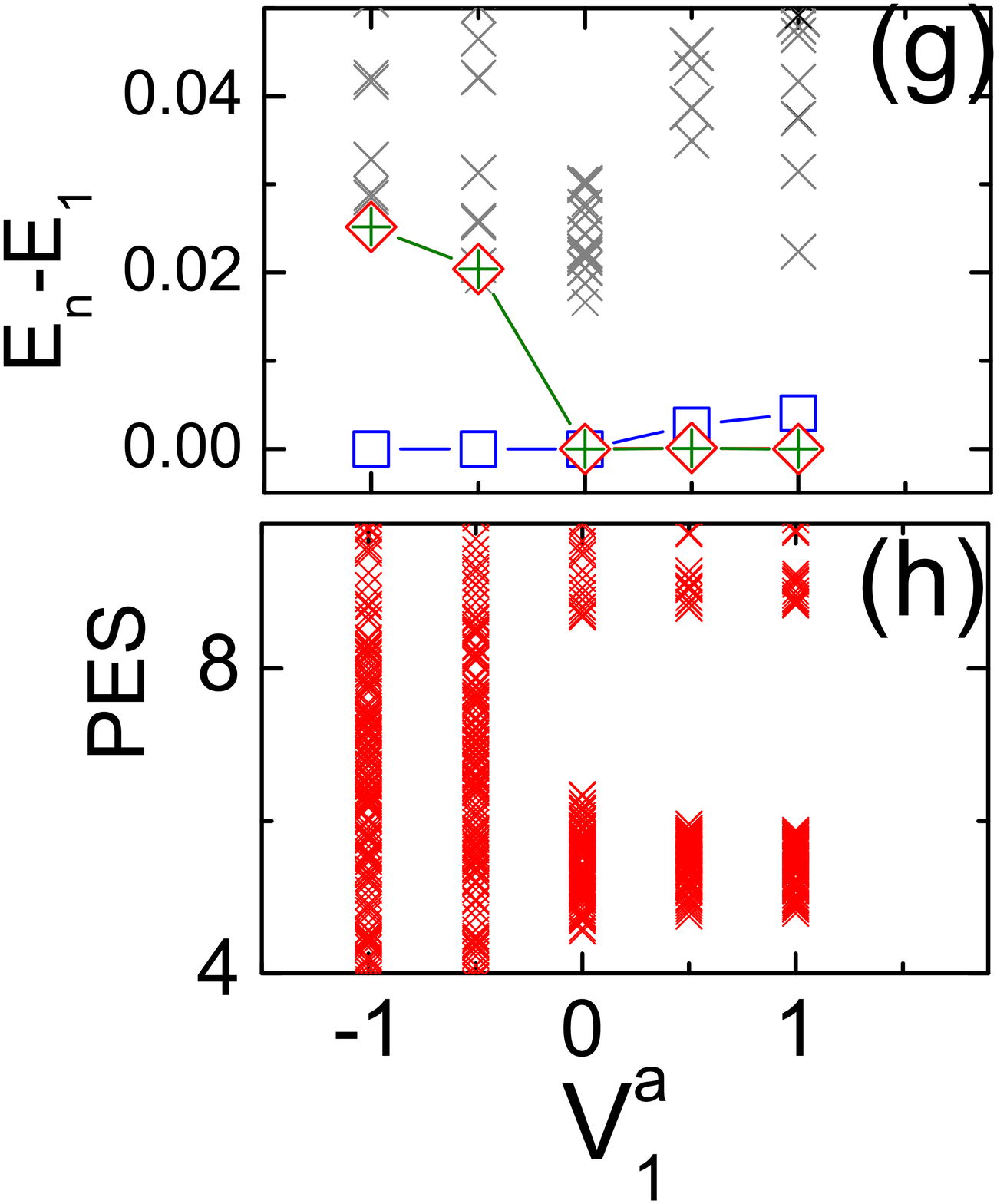}
\end{minipage}
\begin{minipage}{0.208\textwidth}
\includegraphics[width=1.5in]{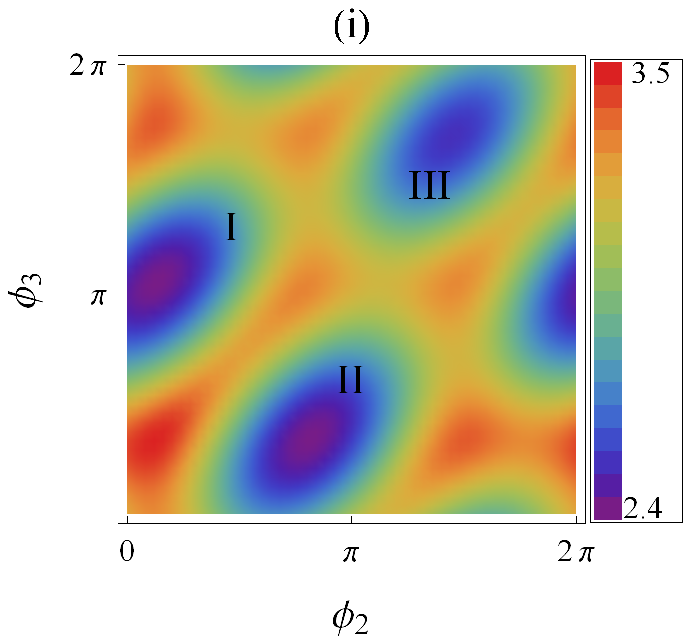}
\end{minipage}%
\begin{minipage}{0.205\textwidth}
\includegraphics[width=1.5in]{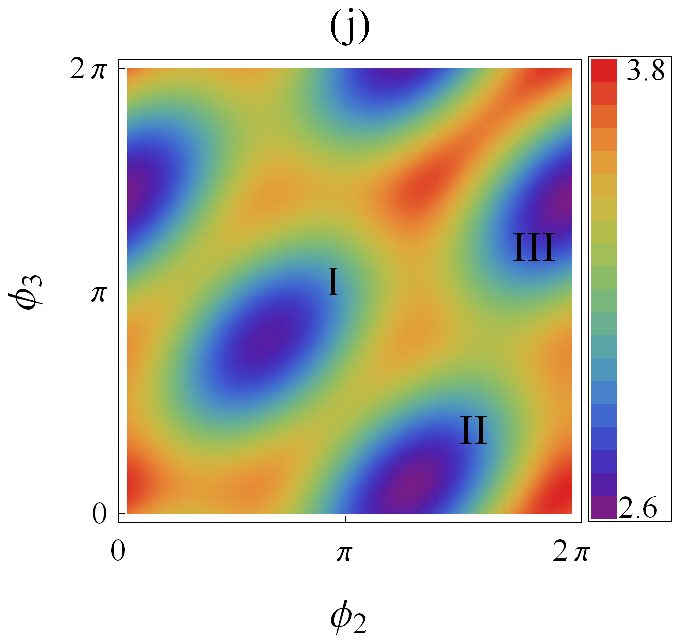}
\end{minipage}
\begin{minipage}{0.205\textwidth}
\includegraphics[width=1.5in]{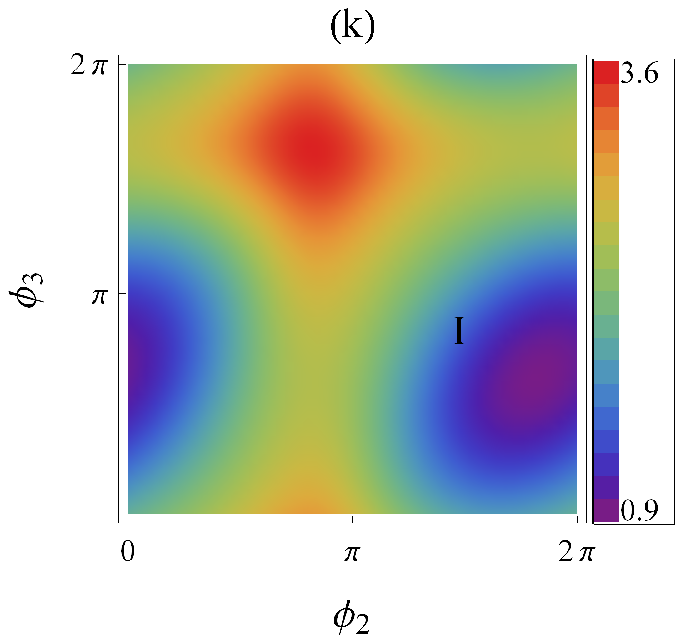}
\end{minipage}
\begin{minipage}{0.205\textwidth}
\includegraphics[width=1.5in]{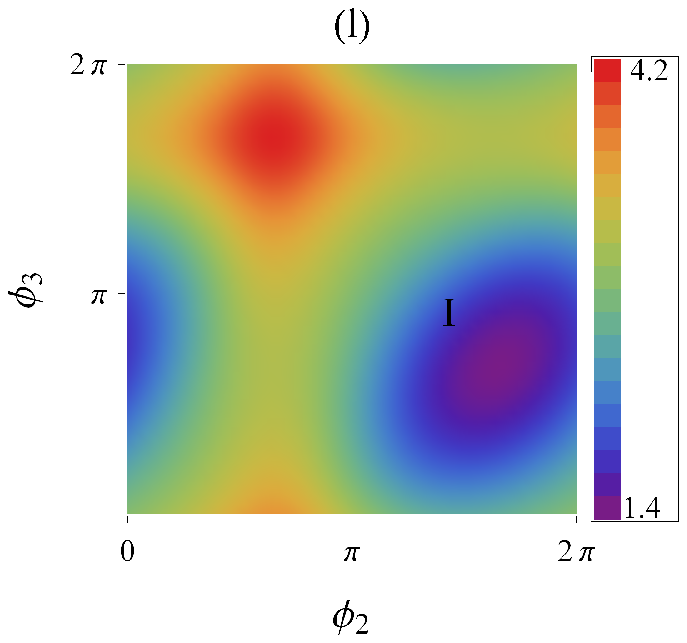}
\end{minipage}
\end{center}
\caption{(Color online) (a-f) Disorder effect on MESs on $2\times 4\times 4$ HC lattice filled with $8$ hard-core bosons.
(a) Energy spectrum (two lowest eigenvalues are labeled by diamond and cross) and (b) Particle entanglement spectrum (PES) for tracing out $5$ particles.
There are $352$ states below the PES gap for $W<0.6$,
in good agreement with the counting of quasihole excitations in FQH state.
Contour plot of entropy for (c) $W=0.1$; (d) $W=0.4$; (e) $W=0.8$; (f) $W=1.0$.
(g-l) Anisotropic interaction effect on MESs on $2\times 4\times 6$ HC lattice filled with $8$ fermions.
(g) Energy spectrum (three lowest eigenvalues are labeled by blue square, red diamond and green cross) and
(h) PES for tracing out $6$ particles. There are $228$ states below the PES gap for $V^{a}_1\geq 0$.
Contour plot of entropy for (i) $V^{a}_1=0.5$; (j) $V^{a}_1=0.0$; (k) $V^{a}_1=-0.5$; (l) $V^{a}_1=-1.0$. Bipartition are all along cut-I direction.
}
\label{fig:hc:disorder}
\end{figure*}

\textit{Topological entanglement entropy.---}
For a topological ordered state, one can also identify  a topological term in
the entanglement entropy  since $S=\alpha L -\gamma +\mathcal{O}(L^{-1})$, where $L$ is
the length of the smooth boundary between two subsystems and the TEE
term is quantized as $\gamma=\log \mathcal{D}$ with $\mathcal{D}$ as the total quantum dimension \cite{Kitaev,Levin}.
Recently, it has been shown that TEE of Abelian FQH state can be extracted
through \cite{Grover}, $\gamma_{cal}=S_{max}-S_{min}$,
where $S_{max(min)}$ is the Renyi $n=2$ entanglement entropy
corresponding to maximal (minimal) entangled state.
To check out this relation,
the calculated $\gamma_{cal}=S_{max}-S_{min}$ for different
systems are shown in Table \ref{table:gamma}.
Indeed, the obtained $\gamma_{cal}$ gives a good estimate of
the quantized theoretical value $\gamma_{Th}=2\times \ln\sqrt{m}=\ln m$ for $1/m-$Laughlin state on torus \cite{Lauchli2010b}.
For symmetric system of  $2\times 4\times 4$ lattice, we obtained
a  bigger deviation between $\gamma_{cal}$ and $\gamma_{Th}$, which may result from the
strong coupling among the groundstates in the same momentum sector.
To elucidate the physical difference between minimal and maximal entangled states,
we further show entanglement spectra $\{\lambda^{max(min)}_i\}$
of these states in Fig.~\ref{fig:es:hc} \cite{Haldane2008,Lauchli2010a}.
We find that the spectra of the maximal entangled states $\{\lambda^{max1(2)}_{i} \}$
can be exactly recovered by reducing the density matrix eigenvalues by a factor
$m$  ($m=2$ for $\nu=1/2$ FQH state) for two sets of spectra of MESs and imposing them on top of each other:
$\{\lambda^{min1}_i/m \} \oplus\{\lambda^{min2}_i/m\}$ as shown as cross dots in Fig.~\ref{fig:es:hc}.

\textit{Modular Matrix and Quantum Phase Transition.---}
Topological order is robust in the presence of
any weak local perturbations, which can be used to characterize the topological phase.
Here  we first consider the disorder effect on bosonic  state.
As shown in Fig.~\ref{fig:hc:disorder}(a-b), the energy spectrum
remains two-fold quasi-degenerating protected by a spectrum gap
until a disorder strength  $W \sim  0.8$.
Further calculation of particle entanglement spectrum (PES)
reveals a gap at small $W$ and the number of states below this gap
agrees with the number of quasihole excitations in a FQH state \cite{Bernevig2011}.
This PES gap  disappears at $W\sim 0.6$ signaling the quantum phase transition
from the FQH  phase to a topological trivial state.
As shown in Fig.~\ref{fig:hc:disorder}(c-d),
there are two distinguishable valleys I and II in  entropy for
the states $|\Phi>=c_1|\xi_1>+c_2e^{i\phi}|\xi_2>$,
and the corresponding MESs are always approximately orthogonal to each other.
The modular matries obtained  for an intermediate disorder strength $W=0.4$ are
$\mathcal{S}\approx 0.685
     \left(
       \begin{array}{cc}
         1 & 1.109 \\
         1.109 & -0.980-0.223i \\
       \end{array}
     \right)$ and $\mathcal{U}\approx
            e^{-i\frac{2\pi}{24}1.12}
            \left(
                 \begin{array}{cc}
                   1 & 0 \\
                   0 & 0.208+0.978i \\
                 \end{array}
            \right)$, which remain to be very close to the exact results for
bosonic $\nu=1/2$ Laughlin state.
After the quantum phase transition  at $W=0.8$
as shown in Fig.~\ref{fig:hc:disorder}(e),  there are
still two valleys of MESs near $(c_1,\phi)=(0.851,1.654\pi)$ and $(0.715,0.684\pi)$,
however these two states start to have bigger  overlap   $|<\Xi_1|\Xi_2>|\approx 0.245$.
The corresponding modular matrix $\mathcal{U}$ and $\mathcal{S}$ are
qualitatively different from exact results for $\nu=1/2$ FQH state as:
$\mathcal{S}\approx 0.851
     \left(
       \begin{array}{cc}
         1 & 0.636 \\
         0.636 & e^{1.328\pi} \\
       \end{array}
     \right)$ and $\mathcal{U}\approx
            \left(
                 \begin{array}{cc}
                   1 & 0 \\
                   0 & 1 \\
                 \end{array}
            \right)$.
In particular, the quasi-particle statistics has changed with the
$\mathcal{U}$ matrix becomes unit matrix,
which indicates we are in a topological trivial phase.
At $W=1.0$ shown in Fig.~\ref{fig:hc:disorder}(e),
there is only one valley left corresponding to one MES state in parameter space
indicating the lost of any feature of topological order.

Furthermore, we consider the effect of the anisotropic interaction
for fermionic system by tunning
the interaction $V^{a}_1$ on one NN bond,
while keeping the other two  at unit strength.
Consistent with the geometrical theory of the FQHE \cite{Haldane2011,Haldane2012},
we find that the topological state and its modular
matrix remain to be universal insensitive to the strength
of the additional repulsive interactions with no quantum phase transition.
So we turn to the additional attractive interaction on one bond.
From both energy spectrum and PES, we identify a quantum phase transition
which appears between $V^{a}_1=0$ and $V^{a}_1=-0.5$ as shown in Fig.~\ref{fig:hc:disorder}(g-h).
As shown in Fig.~\ref{fig:hc:disorder}(i-l), in the FQH phase, there are three
minimal entropy valleys in $\phi_2-\phi_3$ parameter space while
we take $c_1=c_2=c_3$, which are the optimized values for
all these systems  to minimize the entanglement entropy.
In FQH phase, the calculated modular matrix is always nearly identical
 to the expected theoretic result for Laughlin state.
Taking $V^{a}_1=0.0$ as an example, from the overlap of the MESs we extract
$\mathcal{S}\approx
\frac{1}{\sqrt{3}}\left(
       \begin{array}{ccc}
        1 & 1 & 1 \\
        1 & e^{i2\pi\times 0.36} & e^{i2\pi\times 0.68} \\
        1 & e^{i2\pi\times 0.65} & e^{i2\pi\times 0.32}
       \end{array}
     \right) $.
After the phase transition occurs (Fig~\ref{fig:hc:disorder}(k-l))
at $V^a_1 \sim -0.5, -1.0$,
we can only locate one minimal entropy valley
in $\phi_2-\phi_3$ parameter space, which demonstrates the
disappearance of the FQH phase.

\textit{Summary and discussion.---}
We study the structure of MESs in the space of the groundstate manifold obtained from ED calculations.
By calculating the overlap between different MESs, we obtain
modular matrices for different FQH systems.
The obtained $\mathcal{S}$ and $\mathcal{U}$ matrices faithfully represent the
quasiparticle dimension and fractional statistics for systems
with anisotropic interactions and random disorder scattering
until a quantum phase transition takes place.

\textit{Acknowledgements.} We thank Shoushu Gong for discussions.
This work is supported by the US DOE Office of Basic Energy Sciences under Grant
No. DE-FG02-06ER46305 (DNS) and NSF under grants DMR-0906816 (WZ) and the Princeton MRSEC Grant
DMR-0819860 (FDMH).  DNS also acknowledges the travel support by the Princeton MRSEC.

\clearpage

\begin{appendices}

\section{Supplemental material for:``Minimal Entangled States and Modular Matrix for Fractional Quantum Hall Effect in Topological Flat Bands''}
In the main test, we focus on the topological flat-band (TFB) model on honeycomb lattice and
extract the modular matrix and related quasiparticle statistics through locating the
minimal entangled states (MESs). In this supplemental material,
we apply the similar route on checkboard lattice and we focus on
searching the topological order of fractional quantum Hall (FQH) state at filling factor $\nu=1/4$ \cite{YFWang2011}.

The Hamiltonian for checkerboard lattice filled with hard-core bosons\cite{YFWang2011,Sun}:
\begin{eqnarray}
H_{\rm CB}&=& -t\sum_{\langle\mathbf{r}\mathbf{r}^{ \prime}\rangle}
\left[b^{\dagger}_{\mathbf{r}^{ \prime}}b_{\mathbf{r}}\exp\left(i\phi_{\mathbf{r}^{ \prime}\mathbf{r}}\right)+\textrm{H.c.}\right]\nonumber\\
&\pm&t^{\prime}\sum_{\langle\langle\mathbf{r}\mathbf{r}^{
\prime}\rangle\rangle}
\left[b^{\dagger}_{\mathbf{r}^{\prime}}b_{\mathbf{r}}+\textrm{H.c.}\right]
-t^{\prime\prime}\sum_{\langle\langle\langle\mathbf{r}\mathbf{r}^{
\prime}\rangle\rangle\rangle}
\left[b^{\dagger}_{\mathbf{r}^{\prime}}b_{\mathbf{r}}+\textrm{H.c.}\right] \nonumber\\
&+&V_1\sum_{\langle\mathbf{r}\mathbf{r}^{\prime}\rangle}n_{\mathbf{r}}n_{\mathbf{r}^{\prime}}
\end{eqnarray}
where $b^{\dagger}_{\mathbf{r}}$ creates a hard-core boson at site
$\mathbf{r}$,  $n_{\mathbf{r}}=b^{\dagger}_{\mathbf{r}}b_{\mathbf{r}}$
is the boson number operator.
$\langle\dots\rangle$, $\langle\langle\dots\rangle\rangle$ and
$\langle\langle\langle\dots\rangle\rangle\rangle$ denote the nearest-neighbor (NN), the
next-nearest-neighbor (NNN) and the next-next-nearest-neighbor (NNNN) pairs of sites.
We adopt the parameters $t=-1$, $t^{\prime}=1/(2+\sqrt{2})$,
$t^{\prime\prime}=-1/(2+2\sqrt{2})$ and $\phi=\pi/4$, which leads to
a TFB with the flatness ratio about $30$. To stabilize the FQH phase
at filling factor $\nu=1/4$, we set NN interaction $V_1=8.0$
following the previous work\cite{YFWang2011}.

\begin{figure}[b]
 \begin{minipage}{0.55\linewidth}
  \includegraphics[width=2.5in]{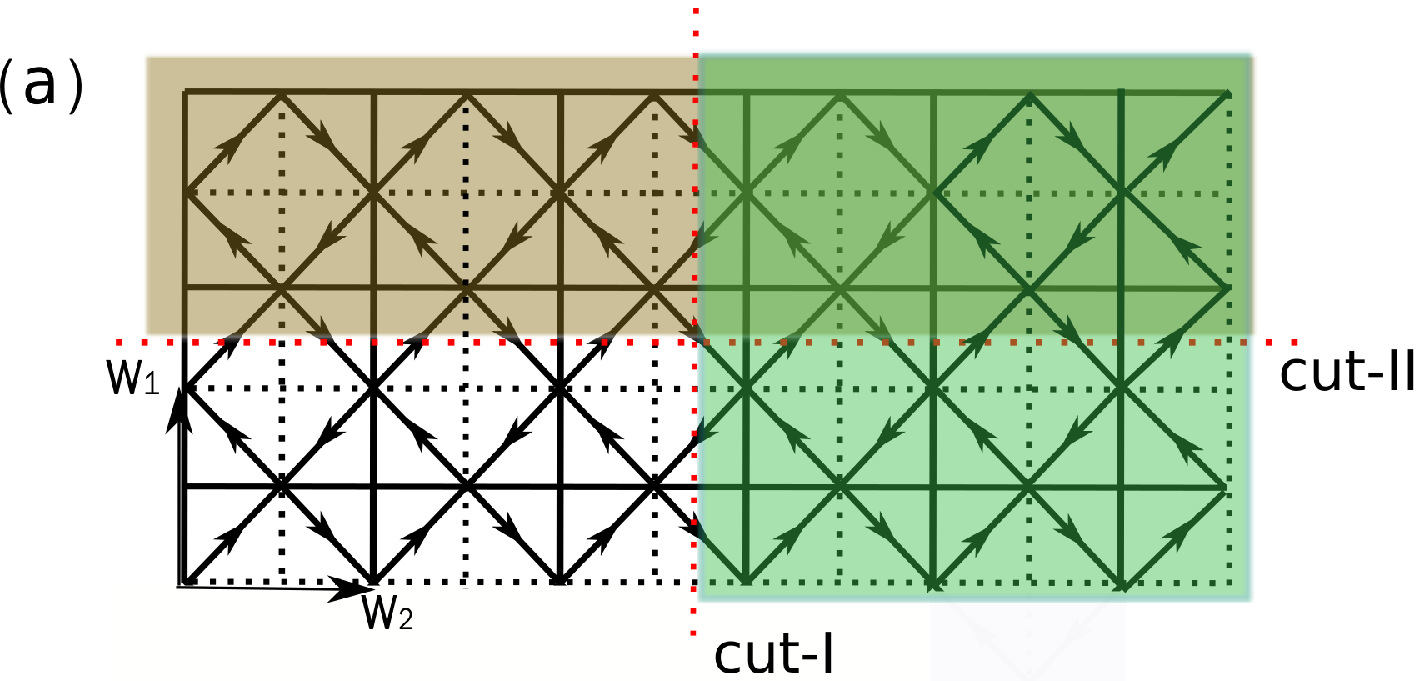}
 \end{minipage}
  \begin{minipage}{0.75\linewidth}
   \includegraphics[width=3.0in]{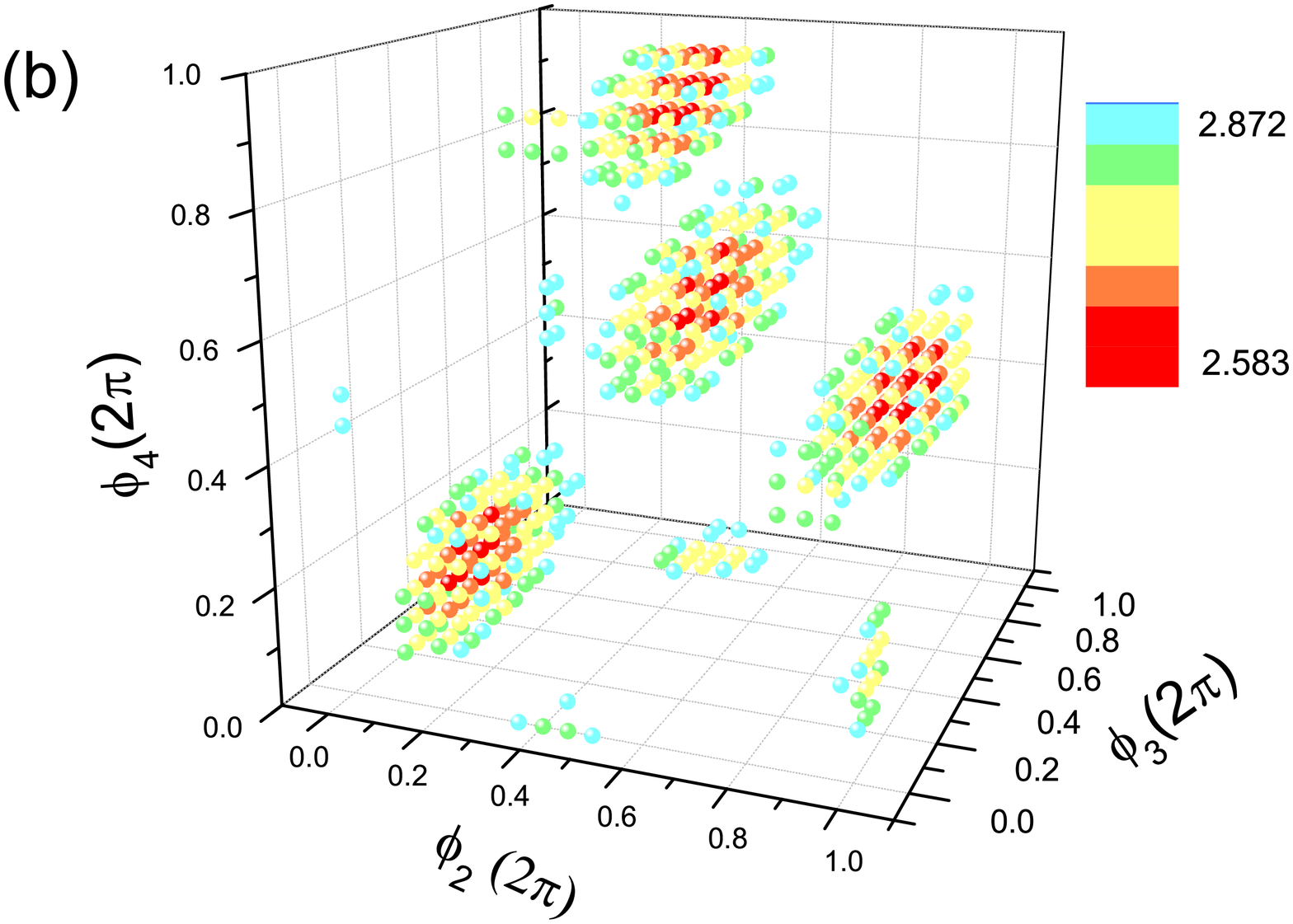}
  \end{minipage}
 \caption{Top: Checkboard lattice with basis vectors $\vec{w}_1,\vec{w}_2$.
 The arrow directions present the signs of the phases $\pm\phi$ in the NN hopping terms.
 The two ways to partition the system along the dashed lines are labeled as cut-I and cut-II, respectively.
 Bottom: The entropy of wavefunction $|\Psi>$ on $2\times 4\times 5$ checkboard lattice with $5$ interacting hard-core bosons by setting $c_1=c_2=c_3=c_4=1/4$.
 Here we only show the entropy smaller than $2.87$. The calculation is for bipartition system along cut-II direction. }
 \label{CB_HB}
\end{figure}

We consider a  $2\times 4\times 5$ checkboard lattice with five hard-core bosons.
In the exact diagonalization study, there are four near degenerating eigenstates which
are separated from higher eigenstates by a finite spectrum gap.
The four ground states $|\xi_i>$ ($i=1,2,3,4$) lie in momentum sector $(k_x,k_y)=(0,0)$,$(1,0)$,$(2,0)$ and $(3,0)$, respectively.
Now we form the general superposition
state from the four quasi-degenerating ground states,
\begin{equation*}
|\Psi>=c_1|\xi_1>+c_2e^{i\phi_2}|\xi_2>+c_3e^{i\phi_3}|\xi_3>+c_4e^{i\phi_4}|\xi_4>
\end{equation*}
where $c_i$ are the real parameters and $\phi_i$ are the relative phase
of the state respectively.
For the bipartition along cut-I,
we find that each groundstate $|\xi_i>$ is indeed the MES
due to four ground states
having different quantum number $k_x$ along the cut-I direction:
\begin{equation}
\label{MES:nu4:x}
  |\Xi^{I}_i> = |\xi_i> ,\,\,\, i=1,2,3,4.
\end{equation}
For the partition along cut-II,
it is found that the MESs appear when the four
ground states are in equal magnitude superposition: $c_1=c_2=c_3=c_4=1/4$.
As shown in Fig.~\ref{CB_HB}, we show the entropy of wavefunction $|\Psi>$
in $\phi_2-\phi_3-\phi_4$ space by setting $c_1=c_2=c_3=c_4=1/4$.
The color of dots represents the magnitude of entropy. For simplicity,
we just show the points with entropy smaller than $2.872$.
It is clear that there exist four valleys in $\phi_2-\phi_3-\phi_4$ space.
The four valleys corresponding to four independent MESs as:
\begin{equation}
\label{MES:nu4:y}
  |\Xi^{II}_i> = (|\xi_1> +e^{i\phi_2(i)}|\xi_2> + e^{i\phi_3(i)}|\xi_3>+ e^{i\phi_4(i)}|\xi_4>)/2 \,
\end{equation}
with state index $i=1,2,3,4$ and
$(\phi_2(1),\phi_3(1),\phi_4(1))=(0.16\pi,0.70\pi,0.26\pi)$,
$(\phi_2(2),\phi_3(2),\phi_4(2))=(0.68\pi,1.72\pi,1.76\pi)$,
$(\phi_2(3),\phi_3(3),\phi_4(3))=(1.16\pi,0.70\pi,1.26\pi)$,
$(\phi_2(4),\phi_3(4),\phi_4(4))=(1.68\pi,1.72\pi,0.76\pi)$,
corresponding to minimum entropies $S=2.592,2.583,2.592,2.583$, respectively.
The four MESs are nearly orthogonal to each other:
$|<\Xi^{II}_1|\Xi^{II}_{2,4}>|\approx 0.089$,
$|<\Xi^{II}_1|\Xi^{II}_3>|\approx 0.0$,
$|<\Xi^{II}_2|\Xi^{II}_4>|\approx 0.0$,
$|<\Xi^{II}_3|\Xi^{II}_{2,4}>|\approx 0.089$,
which forms orthogonal basis states for modular transformation.

As described in the main text, modular matrix can be obtained through
the overlap between MESs along two partition direction: $<\Xi^{II}|\Xi^{I}>=\mathcal{S}$ \cite{YZhang2012}.
Using Eq.\ref{MES:nu4:x} and Eq.\ref{MES:nu4:y}, we obtain,
\begin{equation*}
\mathcal{S}\approx
\frac{1}{2}\left(
       \begin{array}{cccc}
        1 & 1 & 1 & 1\\
        1 & -i & -1 & i\\
        1 & -1 & 1 & -1\\
        1 & i & -1 & -i
       \end{array}
     \right)+10^{-2}e^{i0.49\pi}\left(
            \begin{array}{cccc}
             0 & 0 & 0 & 0\\
             0 & 1 & 1 & 0\\
             0 & 0 & 0 & 0\\
             0 & 1 & 1 & 0
            \end{array}
          \right)
\end{equation*}
, which is nearly the same as $Z_4$ statistics prediction \cite{ZHWang}
up to $10^{-2}$ correction:
\begin{equation*}
\mathcal{S}=
\frac{1}{2}\left(
       \begin{array}{cccc}
        1 & 1 & 1 & 1\\
        1 & -i & -1 & i\\
        1 & -1 & 1 & -1\\
        1 & i & -1 & -i
       \end{array}
     \right)
\end{equation*}
The modular matrix clearly demonstrates topological order of FQH states at $\nu=1/4$.
For example,
from $\mathcal{S}$ we determine:
(i)There are $4$ type quasiparticles in the system labeled by the charges $ae/4$, where $a=0,1,2,3$;
(ii)The quantum dimension of quasiparticles are all $d_i=\mathcal{S}_{i0}/\mathcal{S}_{00} \approx 1,i=0,1,2,3$ thus
the total quantum dimension $D=\sqrt{\sum_id_i^2}=2$;
(iii)The $Z_4$ fusion rule: $a\times b =c$, where $c=Mod(a+b,4),a,b\in 0,1,2,3$ \cite{ZHWang}.

\end{appendices}

\end{document}